\documentclass[]{tCPH2e}

\newcommand{\gsim}{\mbox{\raisebox{-.9ex}{~$\stackrel{\mbox{$>$}}{\sim}$~}}}

\begin{document}

\doi{10.1080/0010751YYxxxxxxxx}
 \issn{1366-5812}
\issnp{0010-7514}

\jvol{00} \jnum{00} \jyear{2009} \jmonth{June}

\markboth{Konstantinos Dimopoulos%
}{Contemporary Physics}

\articletype{\ }

\title{Where galaxies really come from}

\author{Konstantinos Dimopoulos$^{a}$$^{\ast}$\thanks{$^\ast$
Email: k.dimopoulos1@lancaster.ac.uk}\\\vspace{6pt}  
$^{a}${\em{Physics Department, Lancaster University, Lancaster LA1 4YB, UK}}
\\\vspace{6pt}\received{v1 
released April 2009} }

\maketitle

\begin{abstract}
The fundamental paradox of the incompatibility of the observed large-scale 
uniformity of the Universe with the fact that the age of the Universe is finite
is overcome by the introduction of an initial a period of superluminal 
expansion of space, called cosmic inflation. Inflation can also produce the 
small deviations from uniformity needed for the formation of structures in the 
Universe such as galaxies. This is achieved by the conjunction of inflation 
with the quantum vacuum, through the so-called particle production process. 
This mechanism is explained and linked with Hawking radiation of black holes. 
The nature of the particles involved is discussed and the case of using massive
vector boson fields instead of scalar fields is presented, with emphasis on its
distinct observational signatures. Finally, a particular implementation of 
these ideas is included, which can link the formation of galaxies, the standard
model vector bosons and the observed galactic magnetic fields.
\bigskip

\bigskip
\bigskip

\end{abstract}


\section{Introduction}

\subsection{The Hot Big Bang}

\begin{figure}
\begin{center}
\begin{minipage}{100mm}
{\resizebox*{9cm}{!}{\includegraphics{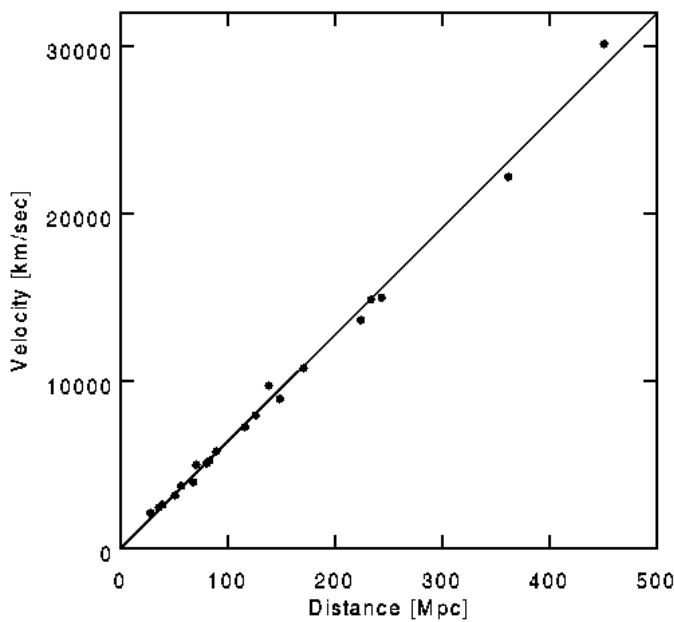}}}%
\caption{%
Observational support for the Hubble Law, which shows the linear increase 
of the apparent recession velocity of galaxies with distance.}%
\label{hubble}
\end{minipage}
\end{center}
\end{figure}

There are a few facts about the Universe we live in that we have now become
certain of. One such fact is that our Universe is expanding.
In 1929 Edwin Hubble discovered that distant galaxies appear to be receding 
from us with velocity proportional to their distance (see Fig.~\ref{hubble}): 
$$
v=H_0r
$$
where $H_0$ is the famous Hubble constant, whose value is 
\mbox{$H_0\simeq 70.5\pm 1.3\,$km\ sec$^{-1}$\ Mpc$^{-1}$} \cite{wmap5}. 
This observation marks the discovery that we live in an expanding 
Universe. Because the above relation is linear it can be deduced that there is
no centre for this expansion (called Hubble expansion); distant galaxies are 
receding away from any observer, wherever placed, according to the above 
relation. The Universe expansion, as it is perceived today, does not correspond
to galaxies travelling through space away from each-other but it is due to
an expansion of space itself, between the galaxies, which is responsible for 
galaxies being pulled away from each-other.\footnote{Note, however, that 
systems which are strongly bound by forces are not pulled apart by the Universe
expansion. This is why galaxies themselves or atoms are not growing in size.}
Hence, according to Hubble Law above, the expansion of the Universe is 
self-similar; the three-dimensional equivalent of a photograph magnification.

\begin{figure}
\begin{center}
\begin{minipage}{100mm}
{
\hspace{-3cm}\resizebox*{16cm}{!}{\includegraphics{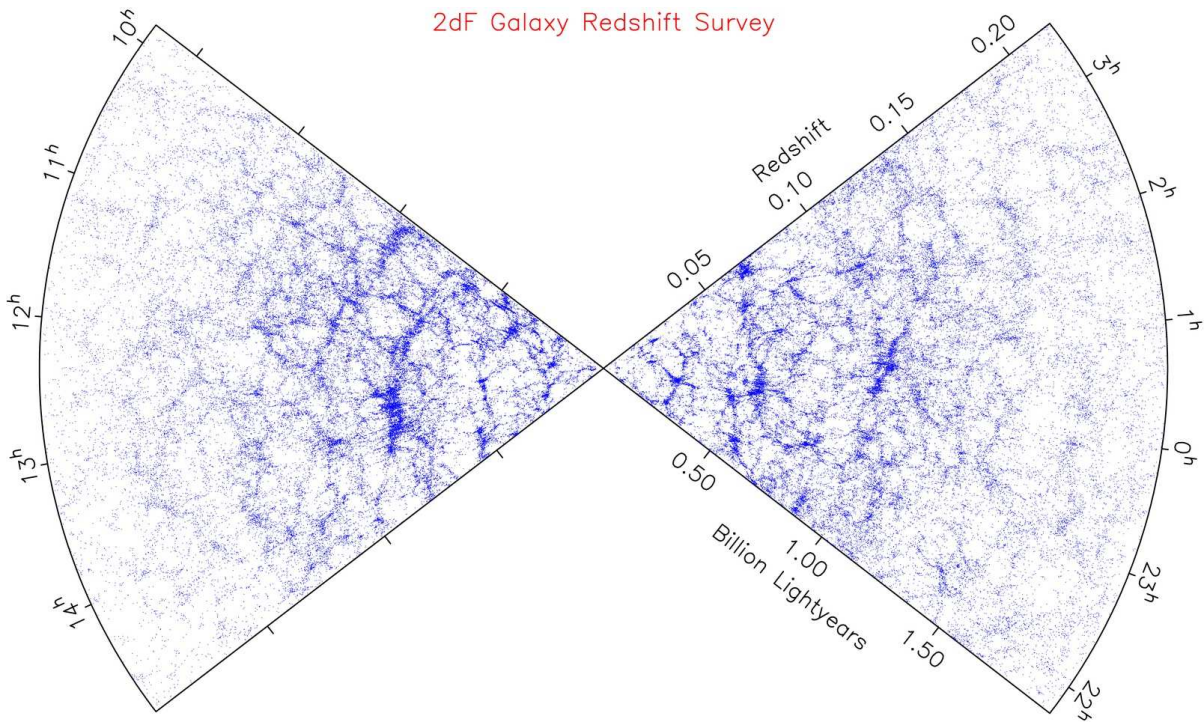}}}%
\caption{%
Image from the 2dF Galaxy Redshift Survey \cite{2dF}. 
Each dot corresponds to a galaxy.
One can clearly see galaxy clusters and superclusters, which form a ``bubbly''
structure with characteristic size $\sim 100\,$Mpc$\,\simeq 0.3\,$Glyrs 
(billion lightyears). One does not see larger structures (e.g. several Glyrs)
spanning the field of vision, which shows that structure in the Universe is not
of fractal~form.
}%
\label{2dF}
\end{minipage}
\end{center}
\end{figure}

Such a uniform expansion suggests that the content of the Universe is also 
uniformly distributed. Indeed, galaxy survey observations show that the 
distribution of galactic clusters and superclusters becomes uniform on scales
above 100~Mpc, which is about a few percent of the size of the observable 
Universe (e.g. see Fig.~\ref{2dF}). That 
is to say that structure in the Universe is not of fractal form; one does not 
find bigger structures the larger the distance scale considered. Instead,
the largest structures correspond to bubble and filament like matter 
distributions comprised by galactic superclusters (see Fig.~\ref{2dF}) whose 
characteristic size ($\sim$~100~Mpc) is much less than the cosmological 
horizon scale which marks the limit of our observational capability. 

The above correspond to observational support for the so-called 
{\sl Cosmological Principle}, 
which simply states that, on large scales, ``The Universe is 
homogeneous and isotropic'', i.e. invariant under translations and rotations.
This means that there are no special points or directions in the Universe, no 
centre, no edges and no axis of rotation. Using the Cosmological Principle it
has been possible to study the global dynamics of the Universe as dictated by
Einstein's general relativity theory.

According to general relativity there is no stable spacetime solution which 
contains uniformly distributed matter and radiation\footnote{%
As is customary in cosmology, by ``matter'' we mean non-relativistic matter
such as galaxies and by ``radiation'' we mean relativistic matter such as 
photons or neutrinos.} and remains static. 
Hence, general relativity demands that
the Universe is either expanding or collapsing. This agrees well with the 
observed Hubble expansion. Now, going backwards in time, we would expect the 
Universe to be contracting and its energy density increasing. 
This means that matter in the Early Universe is predominantly relativistic, 
with a Planckian spectrum of high black-body temperature. Furthermore, the 
Universe content is fully ionised with matter being strongly coupled to 
radiation through Thomson scattering. As the Universe expands, its
density is diluted and its temperature decreases. At some moment, the plasma 
filling the Universe cools down enough to allow neutral atoms to form. When 
this occurs, the scattering of radiation from matter is drastically reduced, 
which means that the Universe content becomes transparent, allowing radiation 
to travel free, similarly to what happens at the photosphere; the visible 
surface of the Sun. At that time radiation is said to ``decouple'' from matter.
This released radiation fills all space and survives until the present time.
Indeed, in 1978 Penzias and Wilson won the Nobel prize because they discovered
this so-called Cosmic Microwave Background (CMB) radiation. 

\begin{figure}
\begin{center}
\begin{minipage}{100mm}
{\resizebox*{10cm}{!}{\includegraphics{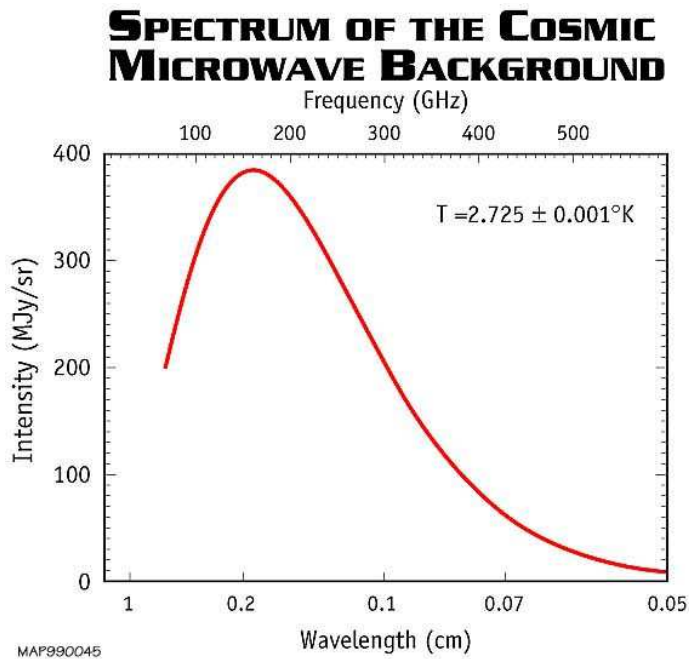}}}%
\caption{%
The observed spectrum of the CMB is an almost perfect black-body corresponding 
to temperature $T_{\rm CMB}=2.725\,$K. The observed spectrum follows the 
theoretical Planck curve so closely that the error bars are impossible to be
depicted at this scale.
}%
\label{cmb}
\end{minipage}
\end{center}
\end{figure}

As shown in Fig,~\ref{cmb}, the CMB has an almost perfect blackbody spectrum 
peaked at microwave 
wavelengths corresponding to temperature of $T_{\rm CMB}=2.725\pm 0.001$~K.
The distribution of the CMB is extremely uniform with minute variations in its 
temperature at the level \mbox{$\delta T/T\approx 10^{-5}$}. Such uniformity
is another piece of supporting evidence for the Cosmological Principle.

The existence of the CMB shows that the Early Universe was hot and
dense and a lot different from the Universe at present. This fact, along with 
the observation that the Universe is expanding, agrees well with 
general relativity which suggests that our expanding Universe has finite age.
This means that, when travelling backwards in time, the reversed expansion is 
such that all distances shrink in a self-similar manner to zero in finite time.
That initial moment of infinite density has been dubbed the Big Bang; the onset
of time itself. 

The latest dynamical estimate for the age of the Universe is
$t_0=13.72\pm 0.12$~Gyrs \cite{wmap5}, which is in good agreement with 
estimates of the ages of the oldest astrophysical objects such as, for example,
globular star clusters \cite{age}.
Hence, cosmologists today operate in the framework of a standard model, which 
considers that the Universe began its existence 14 billion years ago, passed 
through a hot period of ultra-high density, and has continued to cool down 
until the present because 
of the expansion of space. This back-bone history of the Universe is called the
Hot Big Bang and has been observationally confirmed as far back as the time 
when the Universe was no more than a few seconds old.

\subsection{Cosmic Inflation}

Despite the observational support for the Hot Big Bang the fact is that, 
fundamentally, the Cosmological Principle is incompatible with a finite age
for the Universe. This paradox is the so-called {\sl Horizon Problem} and has 
to do with the apparent uniformity of the Universe over distances which are 
causally unconnected. For example, the CMB appears to be correlated at regions 
beyond the causal correlation scale (the so-called particle horizon); it 
appears to be in 
thermal equilibrium and at a preferred reference frame. General relativity 
does not single out a preferred reference frame, which would imply that Doppler
shifts in the frequency of the CMB over causally unconnected regions (with 
different centre of mass frames) would result in $\delta T/T\sim 1$.
In contrast, observations suggest that matter at these regions does lie at the 
same reference frame when the CMB is emitted and the fractional perturbation
of the CMB temperature is much smaller.

\begin{figure}
\begin{center}
\begin{minipage}{100mm}
\subfigure[HORIZON PROBLEM]{
\resizebox*{9cm}{!}{\includegraphics{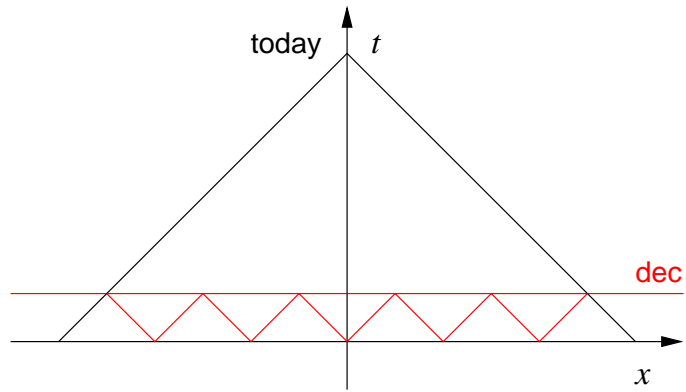}}}\\
\subfigure[INFLATION SOLVES HORIZON PROBLEM]{
\resizebox*{9cm}{!}{\includegraphics{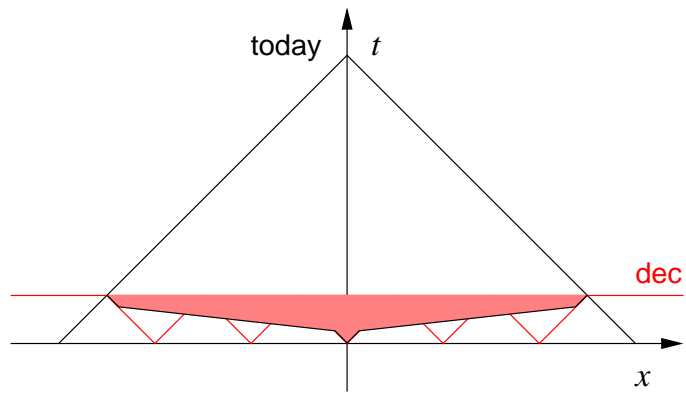}}}%
\caption{%
Space-time diagrams demonstrating the horizon problem and its solution by
cosmic inflation. Slanted lines denote light signals. Here and now is
the tip of our past light-cone, which encloses all events which we can observe 
and can influence us today. The horizontal line corresponds to the epoch of
``decoupling'' when the CMB radiation was emitted. The small triangles in
Figure~(a) depict the light cones (the causally connected regions) at 
decoupling. Figure~(a) shows that our field of vision today includes many 
regions which were unconnected at decoupling. Thus, it is a mystery how the
CMB appears to be in thermal equilibrium at the same temperature in all these 
regions. Figure~(b) depicts how this problem is solved by inserting a period of
superluminal expansion prior to decoupling, which inflates an originally 
causally connected region (inside the light-cone) to size large enough to 
encompass the observable Universe, i.e. the entire field of vision.}%
\label{horizon}
\end{minipage}
\end{center}
\end{figure}

One can understand the problem schematically by considering the spacetime 
diagram in Fig.~\ref{horizon}. The slanted lines depict signals which travel 
with the speed of light, which is the fastest possible propagation through 
space.%
\footnote{Strictly speaking, these lines should be slightly curved because 
of the expansion of the Universe. However, we can ignore this effect if the
expansion is slower than lightspeed.} Along the
time axis the present time $t_0\simeq 1.4\times 10^{10}\,$yrs is depicted at 
the tip of our past light-cone, that is the region of spacetime from which we 
can receive signals and lies below the large slanted lines in the diagram. The 
horizontal line corresponds to the time of the decoupling of radiation from 
matter, when the CMB was emitted at $t_{\rm dec}\simeq 3.5\times 10^5\,$yrs. 
The small triangles correspond to the light-cones at the time of decoupling
and designate the regions which are in causal contact when the CMB was emitted;
at $t_{\rm dec}$ an observer cannot be affected by events occurring beyond 
his/her light-cone. Because $t_0\gg t_{\rm dec}$ our field of vision today 
includes a large number of regions which were beyond causal contact at the time
of the CMB emission\footnote{which reflects how much the range of causal 
correlations has grown since then.}
and yet these
uncorrelated regions appear to be in thermal equilibrium at the same 
temperature.

The most compelling mechanism to date which overcomes the horizon problem is 
the theory of {\sl Cosmic Inflation} originally suggested by 
Guth \cite{guth} and independently by Starobinsky \cite{staro}. 
In a nutshell, cosmic inflation can be 
defined as a brief period of superluminal expansion of space in the early 
Universe. General relativity allows inflation because the latter does not 
correspond to displacement of matter or energy through space with velocity 
faster than lightspeed. Instead, it is space itself which is expanding faster 
than light can travel. Indeed, inflationary expansion is readily obtained by 
general relativity if certain conditions are satisfied.

How does inflation overcome the horizon problem? Well, inflation produces
correlations beyond the causal horizon by expanding an initially causally 
connected region to a size large enough to encompass the observable Universe.
The region which corresponds to the observable Universe is supposed to
lie within the light-cone before the onset of inflation. Being initially 
causally self-correlated it can become uniform by interacting with itself.
After it is inflated to super-horizon distances this original uniformity
sets up uniform initial conditions after inflation, even if the original
region corresponds to many causally-disconnected regions. This is depicted
schematically in Fig.~\ref{horizon}, where during inflation the growth of an 
initially causally correlated region is shown to be faster than lightspeed so 
as to engulf the entire field of vision today.

Hence, we see that, provided it lasts long enough, cosmic inflation can impose
the Cosmological Principle as an initial condition for the Hot Big Bang. 
However, if the cosmological principle was exact 
then there would be no structure in the Universe; no galaxies, no stars with 
planets harbouring intelligent life. All these structures require the 
uniformity of the Universe to be violated. If not, the Universe today would be
filled only with the CMB radiation and a thinned out gas at temperature 
$T_{\rm CMB}$, evenly distributed through space. Our existence demands that 
the Cosmological Principle is only approximately respected. 

It so happens that cosmic inflation can provide also a small violation of the
Cosmological Principle. This is due to quantum effects during inflation, which,
as explained below, can introduce a tiny variation in the density $\rho$ of the
Universe, called the {\sl Primordial Density Perturbation}~$\delta\rho/\rho$.

\begin{figure}
\begin{center}
\begin{minipage}{100mm}
\hspace{-2cm}
{\resizebox*{14cm}{!}{\includegraphics{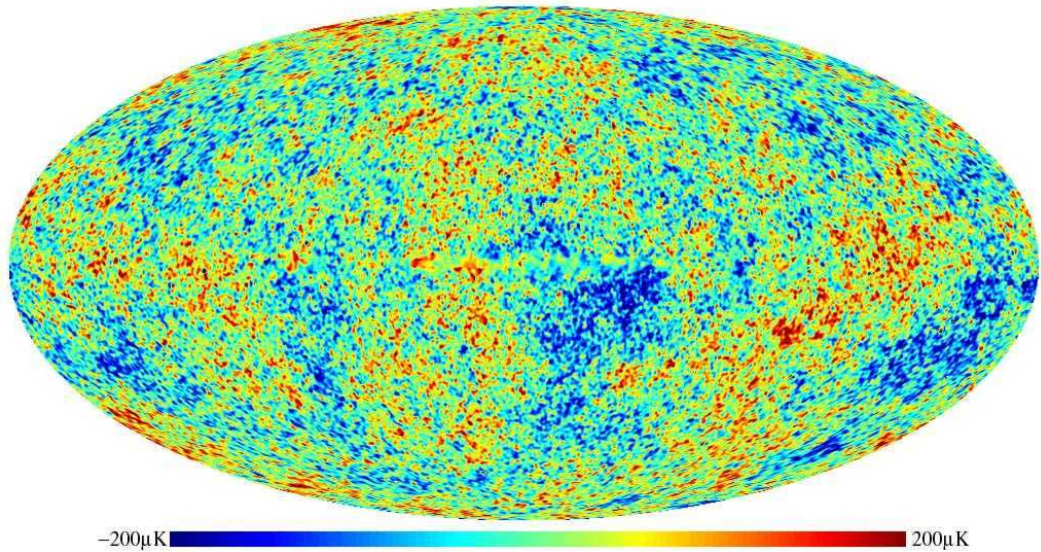}}}\\
\caption{%
Map of the temperature perturbations obtained by the latest observations of the
Wilkinson Microwave Anisotropy Probe satellite (WMAP). Deviations from the 
central value $T_{\rm CMB}=2.7\,$K are colour-coded and depict the range
$|\delta T|\leq 200\,\mu$K, i.e. 
$|\delta T/T_{\rm CMB}|\leq 7.4\times 10^{-5}$.
}%
\label{wmap}
\end{minipage}
\end{center}
\end{figure}

This density perturbation reflects itself onto the CMB through the so-called
Sachs-Wolfe effect \cite{SW}, which describes how the CMB photons become 
redshifted when
crossing regions of density higher than average (so-called overdensities) 
because they lose energy while struggling to exit from the gravitational 
potential wells of these growing overdensities. Hence, variations in the 
density of the Universe cause perturbations in the apparent temperature of the 
CMB radiation:
$$
\left.\frac{\delta T}{T}\right|_{\rm CMB}
=\frac12\,\frac{\delta\rho}{\rho}\approx 10^{-5}.
$$
Indeed the observations (e.g. see Fig.~\ref{wmap}) suggest that 
$\delta\rho/\rho=(1.977\pm 0.039)\times 10^{-5}$ \cite{wmap5}, which, albeit 
tiny, turns out to be enough to explain the formation of structures in the 
Universe, such as galaxies and galactic clusters. 
Starting from an initial density perturbation,
structure formation proceeds through the process of 
{\sl Gravitational Instability}.
The latter amounts to intensifying, in a runaway manner, the contrast between 
overdense and underdense regions and is based on the fact that overdensities 
grow more massive by attracting matter from surrounding underdensities, 
depleting them even further. Indeed, numerical simulations have shown that
density perturbations of magnitude $\delta\rho/\rho\sim 10^{-5}$ at 
$t_{\rm dec}$ suffice to generate the observed structures given 14 billion 
years of growth. Therefore, it seems that inflation can successfully 
impose both the Cosmological Principle onto the Universe and the deviations 
from it, which are necessary for structure formation. 

\section{Particle Production during Cosmic Inflation}

Inflation produces the primordial density perturbation through a process called
{\sl Particle Production} that arises from considering the superluminal 
expansion of space in conjunction with the notion of the quantum vacuum.

\subsection{Classical and Quantum Vacuum}

We are all familiar with the concept of the Classical Vacuum. A box filled with
vacuum is a box which is totally empty of matter and energy, i.e. the 
energy in the box is \mbox{$E=0$}. However, in quantum theory, this cannot be
true and the box has to have \mbox{$E\neq 0$}. This is due to
{\sl Heisenberg's Uncertainty Principle}, which, for energy, can be expressed 
mathematically as\footnote{In quantum mechanics the uncertainty relation is
expressed between momentum and position rather than energy and time, because
time cannot be promoted into a quantum-mechanical operator. However, quantum 
field theory generalises quantum mechanics in four-dimensions by incorporating 
special relativity. The four-dimensional equivalent of the uncertainty relation
does include energy and time as the fourth component of momentum and position
respectively.}
$$
\Delta E\cdot\Delta t\sim\hbar\,.
$$
The meaning of the above expression is that the energy of a closed system (such
as our box) cannot be precisely determined but it has to fluctuate by an amount
$\Delta E$, which, in a given time period $\Delta t$, has to satisfy the above
constraint. The crucial point is that the reduced Planck's constant%
\footnote{\mbox{$\hbar\equiv h/2\pi$}, where $h$ is Planck's constant.}
$\hbar$ featured above, is small but positive, which means
that, for any finite time interval, $\Delta E$ has to be non-zero. In fact, the
smaller the period of time considered the 
larger the fluctuation of the energy has to be. 

\begin{figure}
\begin{center}
\begin{minipage}{100mm}
{
\hspace{1cm}\resizebox*{9cm}{!}{\includegraphics{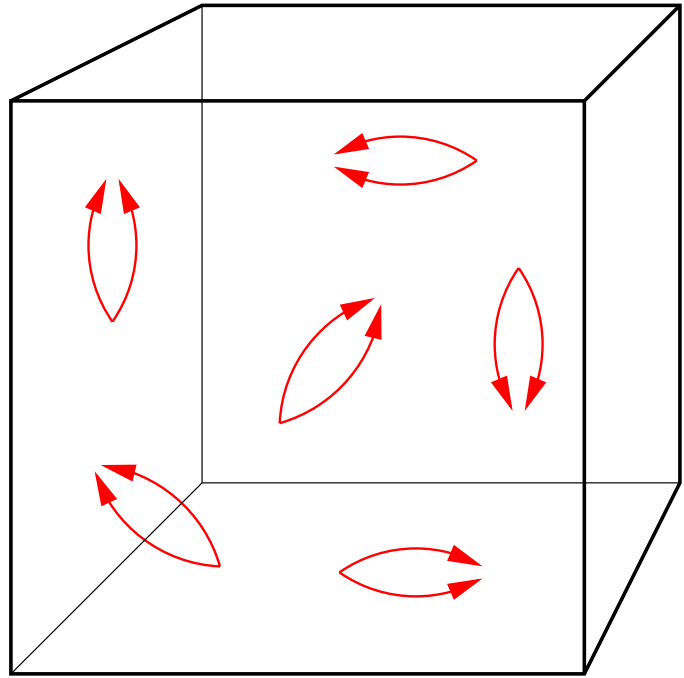}}}%
\caption{%
According to quantum theory, the vacuum is not empty but it is filled with
pairs of virtual particles and anti-particles (whose trajectories are depicted 
here by arrows), which constantly appear from 
(and disappear to) nothing, due to the controlled violation of energy 
conservation enforced by the uncertainly principle. The energy of these
virtual particle pairs is the so-called vacuum energy.
}%
\label{QV}
\end{minipage}
\end{center}
\end{figure}

Hence, the uncertainty 
principle amounts to a controlled violation of energy conservation. This 
violation manifests itself as the brief appearance of pairs of particles and
anti-particles, the reason being that other quantities such as the electric
charge or other quantum numbers, are indeed conserved. Consequently, the 
quantum vacuum is not empty but, instead, it is filled with constantly 
appearing and disappearing pairs of so-called virtual particles and 
anti-particles (see Fig.~\ref{QV}). The energy of these virtual particles is 
the so-called vacuum (or zero-point) energy.

\subsection{The Casimir experiment}

The above may sound like a bunch of flowery ideas, but it so happens that there
is experimental proof of the existence of virtual particles. This is the famous
Casimir experiment, first realised by Casimir and Polder in 1948 
\cite{casimir}. 

\begin{figure}
\begin{center}
\begin{minipage}{100mm}
{
\hspace{1cm}\resizebox*{7cm}{!}{\includegraphics{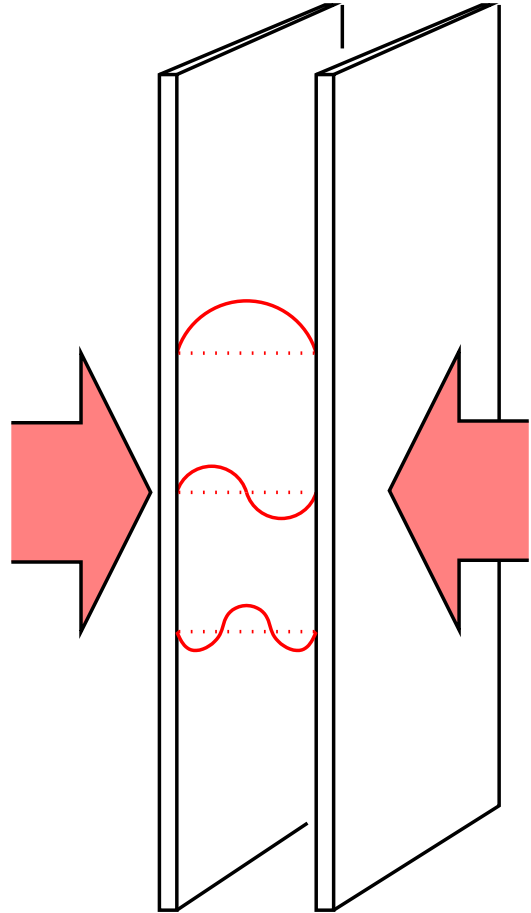}}}%
\caption{%
Schematic representation of the Casimir experiment. Two parallel conducting
plates, not charged and not connected by a circuit, are suspended in vacuum
close to each-other. Since they are conducting they cannot allow a non-zero 
electric field inside them. As a result only a discrete spectrum of virtual 
photons is allowed between them, such that the inter-plate distance 
corresponds to integer numbers of half-wavelengths as shown. Outside the plates
any wavelength is allowed for the virtual photons. This creates an imbalance in
vacuum energy which generates a measurable force on the plates, as depicted by 
the arrows.}%
\label{casimir}
\end{minipage}
\end{center}
\end{figure}

Consider a pair of parallel conducting plates, not electrically charged and not
connected through an electrical circuit. The plates are just standing in empty
space (i.e. in vacuum), close to each-other as shown in Fig.~\ref{casimir}. 
Classically there is no force on 
the plates except from their mutual gravitational attraction, which is 
extremely weak and can be ignored.

Now, let us consider the existence of virtual particles in the space between 
the plates and in particular, the appearance of virtual photons. Photons are 
the quanta of the electromagnetic field. They can be thought of as wave packets
of electromagnetic radiation whose energy $E_\gamma$ is related to their 
wavelength $\lambda$ as
$$
E_\gamma=\frac{2\pi\hbar c}{\lambda}
$$
where $c$ is the speed of light in vacuum. The existence of photons 
has been confirmed by the photo-electric phenomenon,
whose particle interpretation awarded Einstein his Nobel prize in 1921.

Now, since the plates are conducting they cannot allow a non-zero electric 
field inside them. Hence, virtual photons appearing between the plates can only
have a discrete spectrum of energies, corresponding to wavelengths that satisfy
the constraint
$$
N\times\frac{\lambda}{2}=D\,,
$$
where $N$ is a positive integer and $D$ is the distance between the plates.
The above constraint ensures that the amplitude of the electric field remains 
zero at the surface of the conducting plates, as depicted in 
Fig.~\ref{casimir}.

However, the virtual photons appearing outside the plates are not constrained
by the above relation and can have any wavelength. This means that many more
energies of virtual photons are allowed outside the plates compared to the 
discrete spectrum allowed between the plates. Thus, the vacuum energy density 
in the space between the plates is different from the one outside the plates.
This difference (gradient) of vacuum energy gives rise to a force 
$F=-\nabla V$, which has been measured and found to be in precise agreement 
with the predictions of quantum electrodynamics, confirming thereby the 
existence of virtual photons \cite{casimir}.

\subsection{Radiating Black Holes}

Before combining the notion of the quantum vacuum with cosmic inflation we need
to briefly discuss black holes. Black holes are predicted by general relativity
and are thought to exist astrophysically, for example in the centres of 
galaxies. A black hole is an extremely compact object 
with a gravitational field which is locally so intense that anything
approaching close enough can never escape. Indeed, a black hole is surrounded
by an {\sl Event Horizon}, which is the surface on which the escape velocity
from the black hole is equal to the speed of light. This means that one would
need to travel with the speed of light away from the black hole simply to 
remain on the event horizon. Within the horizon the gravitational attraction is
so strong that, even when travelling with lightspeed, one cannot avoid being 
pulled inwards. Since no matter or energy can travel faster than the speed of
light the event horizon forms a boundary on the causal structure of spacetime
in the sense that events within the event horizon cannot affect events outside
it. One can think of the event horizon as a surface which is permeable from
one direction only (from the outside). A classical black hole, being enclosed
within an event horizon, can only absorb matter and energy which is captured 
by its pull from its environment. In-falling matter increases the gravitational
field of the black hole, which in turn increases the radius of its event 
horizon as the gravitational pull further extends its effect in the surrounding
space. Hence, a classical black hole can only increase in mass and size.

\begin{figure}
\begin{center}
\begin{minipage}{100mm}
{
\hspace{1.5cm}\resizebox*{7cm}{!}{\includegraphics{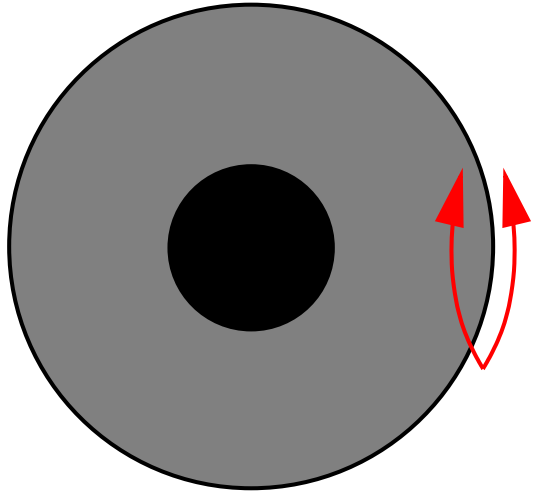}}}%
\caption{%
The appearance of pairs of virtual particles (depicted here by the arrows)
in the vicinity of the event horizon of a black hole, allows the possibility
that one of the particles falls into the event horizon whereas the other one
escapes away.}%
\label{bh}
\end{minipage}
\end{center}
\end{figure}

Such was the understanding of black holes until Hawking studied them in 
conjunction with the quantum vacuum. Hawking considered the appearance of 
virtual particle pairs in the vicinity of the event horizon. Suppose that one
of the particles of the pair falls within the event horizon as shown in 
Fig.~\ref{bh}. The other part of
the pair may follow it, but since it is still outside the event horizon, it has
a non-zero (albeit tiny) probability of escaping from the black hole, whereas,
by definition of the event horizon, the particle within it has no chance of 
escaping. Thus, since pairs of virtual particles are constantly appearing from 
nothing (through the controlled violation of energy conservation by the 
uncertainty principle) in the vicinity of the event horizon, the net effect of
the above possibility is that a tiny fraction of virtual particles does escape 
from the black hole. Such particles cannot meet their counterparts of the 
original pair because the latter have fallen within the event horizon and can 
never escape. Therefore, these particles cannot annihilate with their partners;
their conserved quantum numbers cannot disappear. Consequently, the virtual 
particles which avoided falling within the event horizon (while their partners 
did not) survive and become real particles, which can be detected by distant 
observers \cite{bhpc}.

Stable real particles can survive indefinitely. For stable particles which 
used to be virtual particles that escaped from the event horizon of a black 
hole, this means that $\Delta t$ in the uncertainty relation (that allowed 
their existence) can become very large. Thus, we find 
$\Delta E\sim\hbar/\Delta t\rightarrow 0$, i.e. over large time periods, energy
conservation has to be restored. Since the particles which escaped from the
black hole, can be detected by an observer to have a positive energy 
$E_{\rm out}$, with respect to this distant observer, the pair partners of 
these particles (the ones which did fall into the black hole) must have energy
$E_{\rm in}\approx -E_{\rm out}<0$, where $\Delta E=E_{\rm in}+E_{\rm out}$.
Hence, from the viewpoint of the distant observer, the black hole appears to be
absorbing particles of negative energy which equals in magnitude the positive 
energy of the particles radiated away having escaped from the event horizon. 
Receiving negative energy the black hole reduces in mass and its event horizon
reduces in size, while emitting energy (equivalent to its mass reduction) in 
the form of so-called {\sl Hawking radiation}. Hawking found that the emitted 
radiation has a black-body spectrum corresponding to a characteristic 
temperature, called the {\sl Hawking temperature} $T_H$ \cite{bht}. 
Surprisingly, $T_H$ was found to increase the smaller the black hole becomes, 
i.e. the emission of Hawking radiation intensifies as the black hole shrinks. 
This runaway behaviour ends up in the evaporation of the black hole 
\cite{evap}. Turning this around, the emission
of Hawking radiation from supermassive astrophysical black holes, such as the 
ones in the centres of galaxies, is negligible and is overwhelmed by absorption
of positive energy from their environment, such as surrounding matter or even 
the CMB radiation.

\subsection{Density perturbations from Inflation}

\begin{figure}
\begin{center}
\begin{minipage}{100mm}
{
\hspace{-4cm}\resizebox*{18cm}{!}{\includegraphics{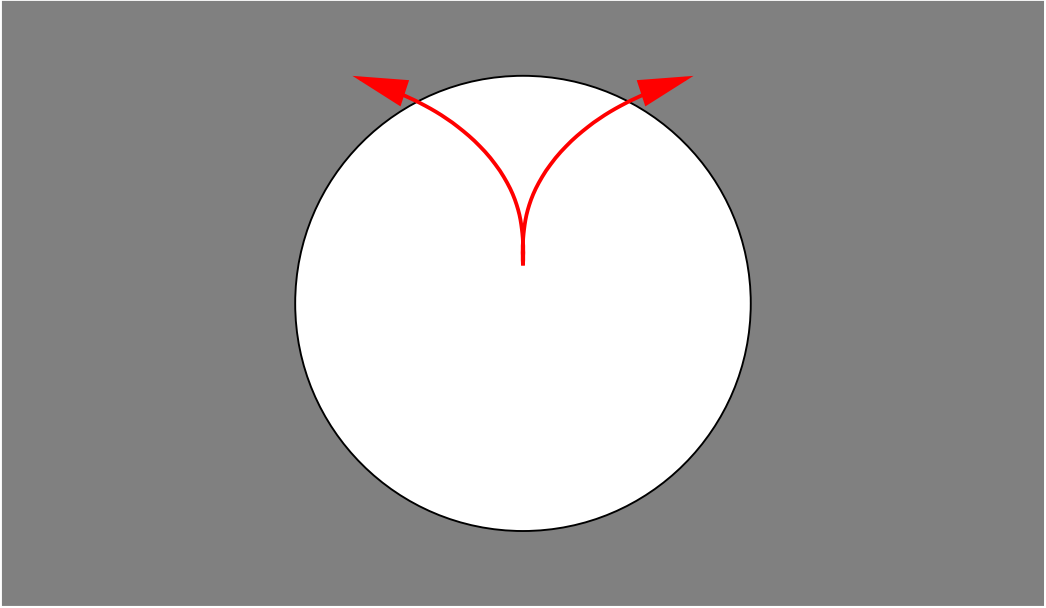}}
\vspace{-3cm}}%
\caption{%
The cosmological horizon during inflation behaves as an event horizon of an
``inside-out'' black-hole in the sense that matter is ``sucked-out'' by the
superluminal expansion. The same can be true for pairs of virtual particles
(depicted here by the arrows).
Indeed, virtual partners, after being pulled outside the horizon, can no more 
find each-other and annihilate, becoming thereby real particles.}%
\label{superlum}
\end{minipage}
\end{center}
\end{figure}

What does black hole evaporation have to do with cosmic inflation? Well, it so 
happens that the cosmological horizon (the range of causal correlations) during
inflation corresponds to an event horizon of an ``inverted'' black hole; a 
black hole ``inside-out'', centred at the observer. This can be understood by
considering that, during inflation, matter inside the horizon is being 
``sucked out'' by the superluminal expansion of space, in analogy with
the fact that nearby matter is being ``sucked in'' by a black hole. In this 
manner the virtual particle pairs of the quantum vacuum can be pulled outside 
the horizon before they have a chance to annihilate as shown in 
Fig.~\ref{superlum}. In this case, being over
superhorizon distance apart, they are beyond causal contact and can never find
each-other to annihilate. Thus, they cease being virtual particles and become 
real particles instead. This is why this process is called 
{\sl Particle Production} \cite{dspc}.

It is important to state here that the existence of an event horizon suffices 
to have particle production. It is the division of the causal structure of 
spacetime, enforced by the event horizon, which gives rise to the particle
production process; the separation of the virtual particle pairs. Thus, the 
event horizon, once it exists, can be seen to emit Hawking radiation without 
the need of a black hole or any concentration of mass.

During inflation, the event horizon is filled with Hawking radiation
(since it corresponds to the outside of the ``inverted'' black hole). Since,
the horizon is centred at the observer and we can put an observer anywhere in
space, this means that all space is filled with Hawking radiation, i.e. 
particle production occurs everywhere once we have superluminal expansion.

Now, in quantum theory, particles correspond to waves with an associated
de Broglie wavelength. For example, a massive particle of mass $m$ is 
characterised by a wavelength $\lambda=2\pi\hbar/mc$. This dual nature of 
particles is hard to visualise, although we are all familiar with many 
``particle-like'' properties of waves, such as the reflection of an incident 
sea wave on a concrete wall. It is also easy to accept that waves, like 
particles, carry energy and momentum, e.g. radio-waves, laser beams or 
tsunami waves. Particles correspond to waves travelling on an otherwise 
smooth sea, which we call a field (the particle is part of the field as the 
sea wave is part of the sea), for example photons travel through the sea of the
electromagnetic field. Virtual particles of the quantum vacuum, therefore, 
correspond to spontaneously arising ripples on the calm sea of their 
corresponding fields. This is why they are also referred to as quantum
fluctuations of these~fields. 

Based on the above, the particle production process during inflation
can be also understood as the stretching of quantum fluctuations by the
superluminal expansion of space, to superhorizon distances. This stretching
transforms the quantum fluctuations into classical perturbations of the fields.
One can imagine them as mountains and valleys, corresponding to classical 
variations of the field in question over superhorizon distances. A variation of
the values of fields gives rise to a variation of their energy density. This,
therefore, is the way that particle production during inflation produces 
variations of the density of the Universe which generate the primordial density
perturbation, that sources the formation of structures in the Universe 
\cite{book}. Schematically, we can write
$$
[\,\Delta E\cdot\Delta t\sim\hbar\,]+{\sf INFLATION}
\;\rightarrow\;
\frac{\delta\phi}{\phi}
\;\rightarrow\;
\frac{\delta\rho}{\rho}
\;\rightarrow\;{\sf GALAXIES}
$$
where $\phi$ is one of the fields in question, whose typical variation during
inflation is given by the Hawking temperature $\delta\phi\simeq T_H$. Thus, 
according to this scenario, all structures in the Universe, 
such as galaxies, stars, planets and ultimately ourselves, originated as 
quantum fluctuations (like the ones we observe in the lab through the Casimir 
experiment) during a period of superluminal expansion of space.

\subsection{Inflation and observations}

\begin{figure}
\begin{center}
\begin{minipage}{100mm}
{
\hspace{.5cm}\resizebox*{9cm}{!}{\includegraphics{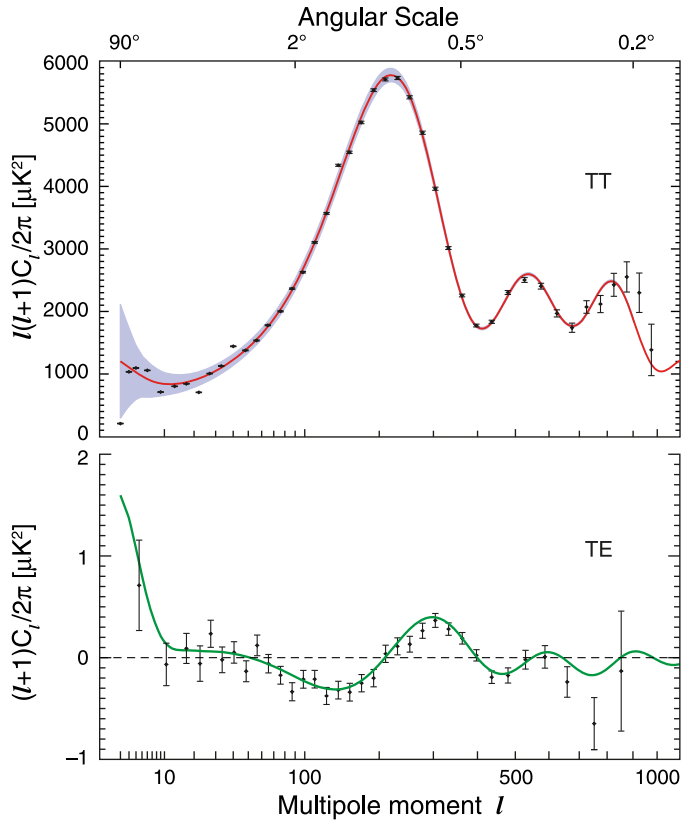}}}%
\caption{%
Angular spectrum of the CMB primordial anisotropy depicting the 
temperature-temperature (TT) and temperature-polarisation (TE) correlations
as obtained from the latest (5-year) observations by the WMAP satellite
\cite{wmap5}.
The solid lines correspond to the predictions of inflation, which are in 
remarkable agreement (especially for the TT plot) with the data.}%
\label{peaks}
\end{minipage}
\end{center}
\end{figure}

Do we have any observational support for this amazing scenario? Indeed we do.
High precision observations of the CMB temperature perturbations, which are 
due to the primordial density perturbation, have revealed a stunning agreement 
with the predictions of inflation. Fig.~\ref{peaks} depicts an analysis of 
these temperature perturbations in spherical harmonics, which follows the 
theoretical prediction of inflation (shown by the solid lines) with remarkable 
consistency \cite{wmap5}. 
It was this kind of analysis that led to the collapse of a rival 
theory for the generation of the density perturbations 
(that of cosmic strings), which did not predict a sequence of peaks in the 
angular spectrum \cite{leandros}
(they are clearly observed in Fig.~\ref{peaks}). 

The simplest and most natural realisations of inflation suggest that the 
Hawking temperature $T_H$ remains roughly constant during the inflationary
phase of the Universe. This means that the variations of the fields, produced
by the particle production process, are of the same amplitude even though they
may correspond to much different length-scales, because the quantum 
fluctuations which exit the horizon early are stretched to much larger 
distances than those which exit the horizon later on. As a result we expect 
that the spectrum of the density perturbations is almost scale-invariant, i.e. 
their amplitude is independent of the length-scale considered. Indeed, 
parameterising the
scale dependence of the perturbation spectrum on the wavenumber~$k$ as 
${\cal P}(k)\propto k^{n_s-1}$, we expect that the spectral index would be
$n_s\approx 1$, where $k=2\pi/\ell$ with $\ell$ being the corresponding 
length-scale. The latest observations suggest that\footnote{%
for negligible tensor contribution.} $n_s=0.960\pm 0.014$ \cite{wmap5}, which
is very close to scale-invariance exactly as inflation predicts. The deviation 
from exact scale invariance was also expected, and has to do with the fact that
the inflationary expansion has to end well before the first second of the 
Universe history.

Another feature of the observed CMB temperature perturbations which agrees well
with inflation is the fact that they appear to be predominantly Gaussian. The 
nature of quantum fluctuations is stochastic (that is to say random) which 
suggests that $\delta\phi$ has a Gaussian bell-shaped distribution centred at
$T_H$. Thus, inflation naturally predicts Gaussian perturbations, as is indeed 
observed.

The most important feature of the observations which supports the inflationary
scenario for the generation of the primordial density perturbation is the 
following. The main peak in the angular power spectrum in Fig.~\ref{peaks}
corresponds to the scale of the cosmological horizon at the time $t_{\rm dec}$ 
of the emission of the CMB. As shown in Fig.~\ref{peaks}, this peak corresponds
to the multipole moment with $l\approx 200$. In the bottom half (the TE part) 
of Fig.~\ref{peaks}, we see that there is an anti-correlation feature at 
multipole moments significantly lower than the
one of the main peak, which correspond to superhorizon scales at the time of 
the emission of the CMB. This means that the density perturbations show 
correlations beyond the causal horizon, a feature that only the superluminal 
expansion of an inflationary phase can explain. 

\section{Density perturbations from Vector Boson Fields}

\subsection{Which fields to use?}

Looking into the particle production process in more depth one has to wonder
which kind of fields are the ones whose quantum fluctuations are responsible
for the generation of the primordial density perturbation. Until recently, all
mechanisms considered the use of scalar fields for the density perturbations.
Scalar fields are hypothetical spin-zero fields, which are characterised by
one degree of freedom, that is a single value at a given point in space.
Scalar fields are ubiquitous in theories beyond the standard model of particle
physics, such as Supersymmetry (the scalar partners of the observed fermion 
fields) or String Theory (the so-called moduli fields, which parametrise the
size and shape of the hypothesised extra dimensions). However, it has to be
stressed that no scalar field has ever been observed. This means that designing
models using unobserved scalar fields undermines their predictability and 
falsifiability, despite the high precision of the recent cosmological
observations. 

The standard model of particle physics does contain one scalar field, which
is the only piece of it that has not been experimentally confirmed. This is 
the famous Higgs field, thought to be responsible for the masses of standard
model particles (which are all the known particles such as electrons, quarks 
and so on). Observing the quanta of this field, the Higgs bosons, is one of the
primary objectives of the Large Hadron Collider (LHC), which is about to become
operational in the coming months in CERN, Geneva. Observation of the Higgs
particles, among other things, will substantiate the claim that scalar fields
do exist and are not merely mathematical constructions. However, what if the
Higgs bosons are not found by the LHC? Do we have an alternative proposal for
the generation of the density perturbations from inflation? Can we form the 
galaxies without scalar fields?

\subsection{The case of Vector Fields}

Recently, a proposal alternative to scalar fields has been put forward, which
uses vector boson fields instead \cite{vecurv}. 
These are spin-one fields, characterised by three degrees of freedom 
corresponding to their magnitude, direction and orientation at every point in 
space. In contrast to scalar fields, vector bosons are indeed observed. The 
standard model of particle physics contains four vector boson fields: one is 
the familiar photon, i.e. the quantum of the electromagnetic field,
and then we have the three massive vector bosons of the weak nuclear force%
\footnote{The weak nuclear force is one of the fundamental interactions and
is responsible for a certain type of radioactivity, called $\beta$-decay,
which transforms neutrons to protons 
by emission of electrons, otherwise known as $\beta$-radiation.},
which are the so-called Z-boson and the W$^{\pm}$-bosons, that have been 
observed in CERN in the '80s.

Vector bosons were not originally considered as candidates for the particle
production process in inflation because of two generic obstacles, which were
thought to inhibit their suitability for this role. The major obstacle has to
do with the inherent anisotropic nature of vector fields. Inflation would 
homogenise a vector field as it does with everything else in the Universe 
when it solves the horizon problem.
Now, if a vector field is to generate or even affect the primordial density 
perturbation it needs to dominate (or nearly dominate) the content of the 
Universe at some time so that it can affect the Universe expansion rate.
However, a uniform vector field (homogenised by inflation) is, in general,
anisotropic because it picks up a preferred direction in space. If such an 
anisotropic contribution to the density of the Universe became dominant it 
would result in anisotropic expansion which would, ultimately, be in conflict
with the predominant isotropy of the uniform CMB radiation. However, it can
be shown that this problem can be overcome if the vector field is undergoing
rapid coherent oscillations when it dominates the Universe \cite{vecurv}.

The equation of motion for a homogeneous massive (and Abelian) vector field in 
the expanding Universe is
$$
\ddot A+H\dot A+m^2A=0\,,
$$
where the dot denotes time derivative, $m$ is the mass of the field and $H$ is
the fractional rate of the Universe expansion; the so-called Hubble parameter,
whose value today is the Hubble constant $H_0$ featured in the Hubble Law. 
Originally the vector field is taken to be subdominant, i.e. its contribution
to the density budget of the Universe is negligible. After inflation the 
Universe expansion rate is gradually decreasing so that the value of the 
Hubble parameter $H$ diminishes. Eventually, we have $H<m$, which means that
one can ignore the second term in the above equation of motion (the so-called 
friction term) rendering the equation similar to the one of a harmonic 
oscillator. Hence, the homogeneous massive vector field eventually undergoes 
harmonic oscillations, with frequency $m$, much larger than the inverse of the
characteristic timescale of the Universe expansion, given by $H$. These 
harmonic oscillations rapidly alternate the orientation of the vector field.
As a result, the oscillating massive homogeneous vector field features no net 
direction and behaves like an isotropic fluid. Therefore, it can dominate the 
Universe content without generating an excessive large-scale anisotropy 
\cite{vecurv}, which would otherwise be in conflict with the~CMB.

The second obstacle to the use of vector bosons for the particle production
process is more subtle. In the case of scalar fields, particle production 
during inflation requires the mass of the virtual particles to be small 
$m<T_H$, otherwise they annihilate before being pulled over superhorizon 
distances by the superluminal expansion. Hence, scalar fields need to be light 
to undergo particle production during inflation. If the same argument is 
applied to vector fields then a problem arises: light vector boson fields are 
approximately conformally invariant, which inhibits their particle production.
This can be understood as follows.

A massless vector field (e.g. the photon) is conformally invariant. This 
suggests that it is 
unaffected by the expansion of the Universe, because it perceives this 
expansion as a conformal transformation to which it is insensitive. 
In terms of its virtual particles, this means that they are not pulled outside
the horizon during inflation by the superluminal expansion. Hence, the quantum 
fluctuations of a massless vector field are not stretched by the expansion to
become classical perturbations. Consequently, a light vector field is 
approximately 
conformally invariant, so one would expect the particle production process to 
be suppressed. To overcome this obstacle an explicit breakdown of the vector 
field conformality is required. Fortunately, there are numerous mechanisms in
the literature which achieve this\footnote{%
These mechanisms were developed for the generation of primordial magnetic 
fields during inflation (for a review see \cite{pmfrev}).}, 
but the disadvantage is that particle 
production is model-dependent; it depends on the mechanism considered.
On the positive side though, this very fact may allow us to discriminate 
between mechanisms, through comparison with observations.

\subsection{Distinct observational signatures}

\begin{figure}
\begin{center}
\begin{minipage}{100mm}
{\resizebox*{10cm}{!}{\psfig{file=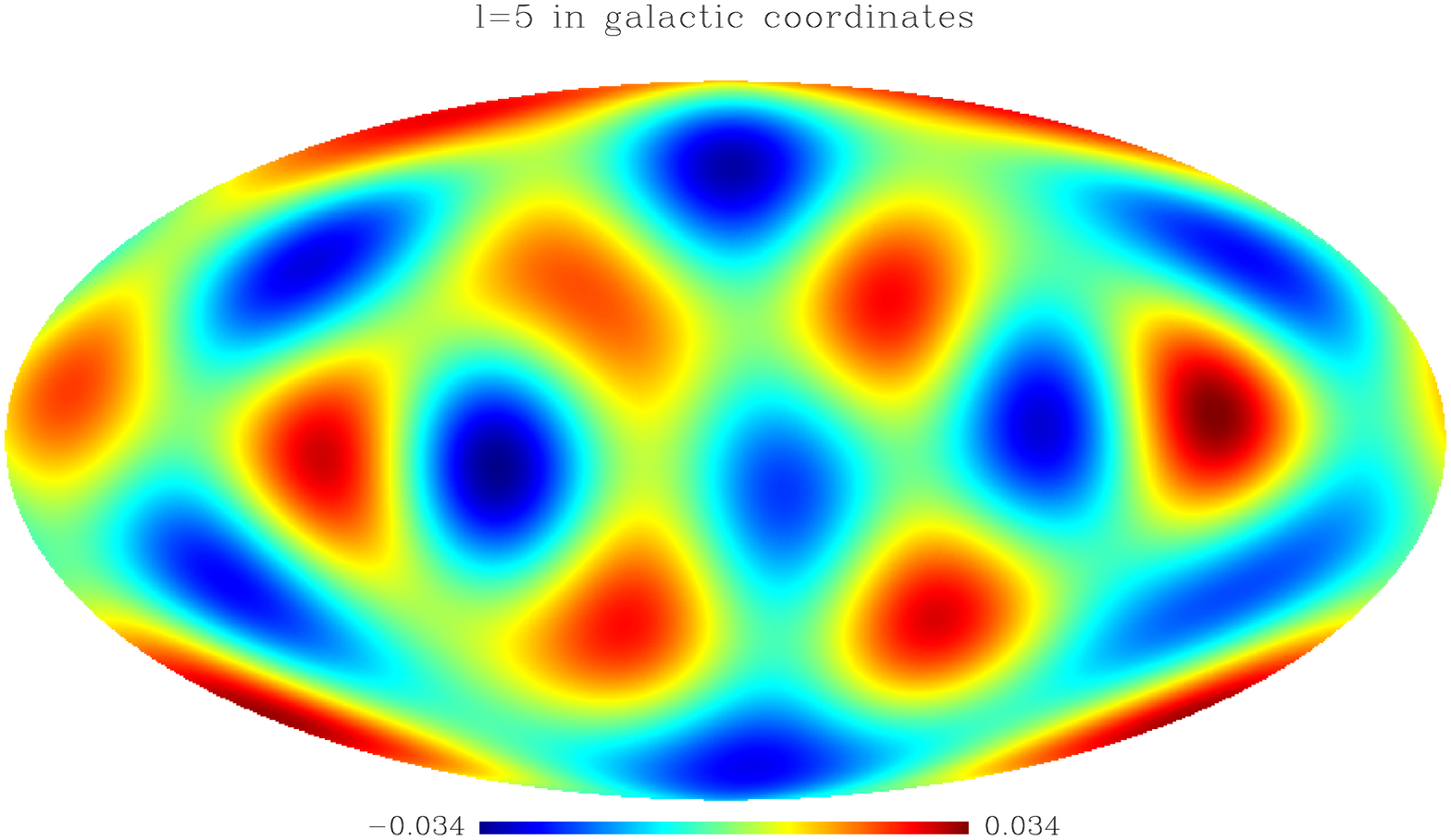,angle=90,width=9cm}}}%
\hspace{1cm}
{\resizebox*{10cm}{!}{\psfig{file=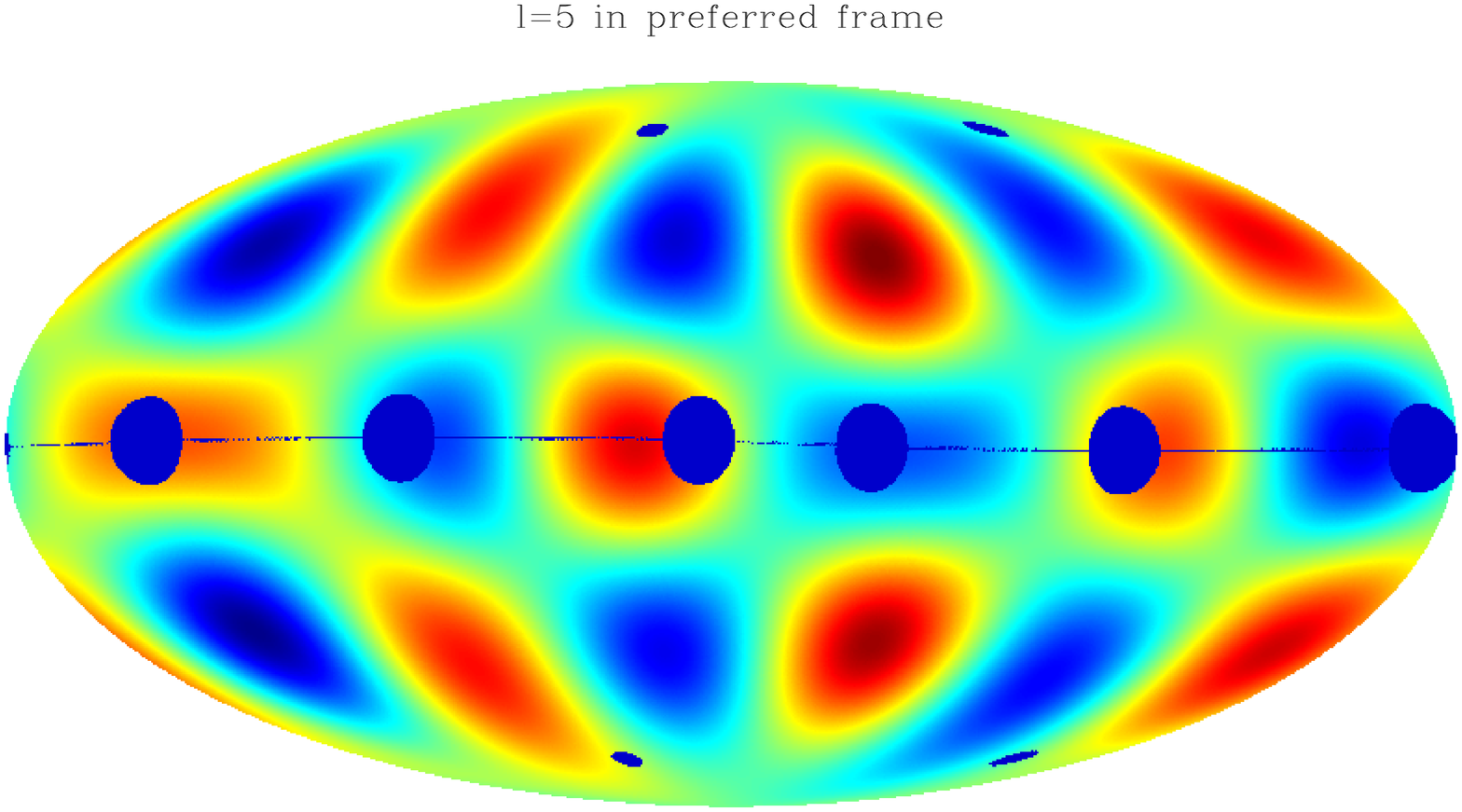,angle=90,width=9cm}}}%
\caption{%
Depiction of the ``Axis of Evil'' observation \cite{AoE}. The upper graph shows
the 5th multipole of the CMB temperature perturbations plotted in the usual
galactic coordinates (as is Fig.~\ref{wmap}). In the lower graph the 5th 
multipole is plotted along a preferred direction, resulting in the emergence of
a clearly seen pattern.}%
\label{AoE}
\end{minipage}
\end{center}
\end{figure}

There are a number of observational characteristics which may signify the
contribution of a vector field to the primordial density perturbation. For 
example, in the above described mechanism which employed an oscillating
massive vector field, there is a chance to produce a weak large-scale 
anisotropy in the Universe. The reason is that the coherent oscillations of
the massive vector field are not exactly harmonic because their amplitude
is gradually decreasing as their energy density is depleted by the Universe 
expansion. Because of this, the oscillating vector field is not exactly 
isotropic, which means that, if it dominates the Universe, it may give rise
to a weak large-scale anisotropy. It so happens that there is recent 
tantalising evidence that such a weak large-scale anisotropy does exist in the 
CMB. Indeed, there is a highly improbable correlation found between the 
quadrupole and octupole moments in the CMB angular spectrum, which reveals the 
existence or a preferred direction, called the ``Axis of Evil'' \cite{AoE}. 
Plotted along this direction
(instead of the usual galactic coordinates) low multipoles in the CMB produce
a clear pattern, as shown in Fig.~\ref{AoE}. An oscillating vector field may 
well explain this feature, which is impossible to account for under the 
traditional scalar field mechanisms.

\begin{figure}
\begin{center}
\begin{minipage}{100mm}
\mbox{\hspace{-2.8cm}%
{
\resizebox*{5cm}{!}{\includegraphics{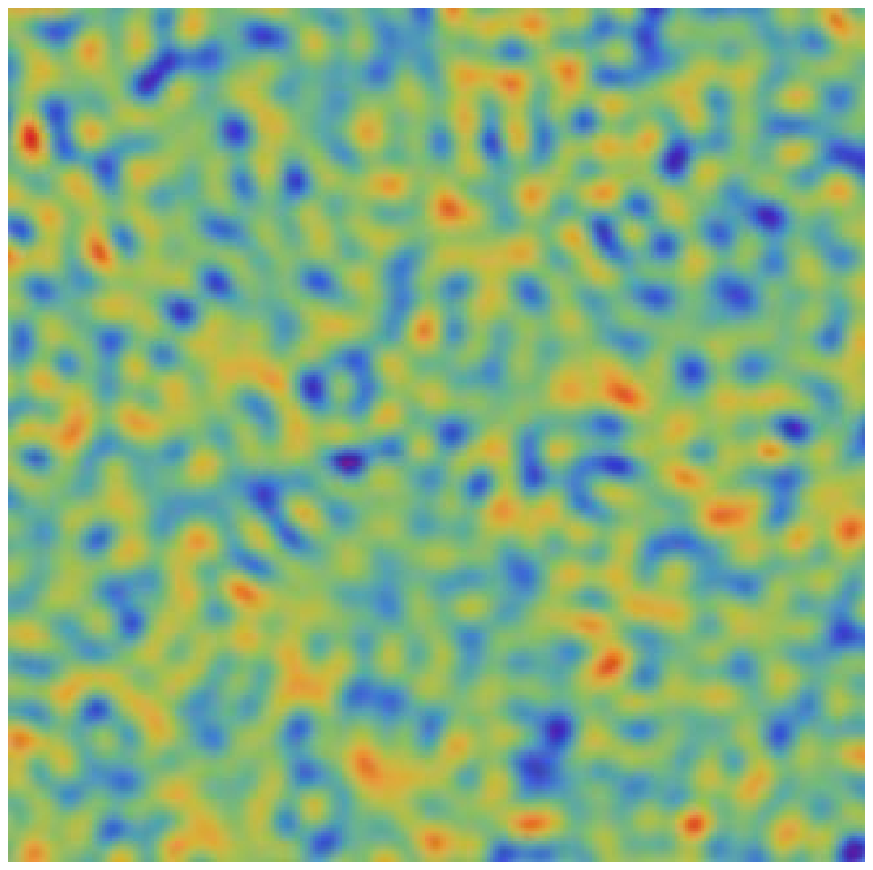}}}
{
\resizebox*{5cm}{!}{\includegraphics{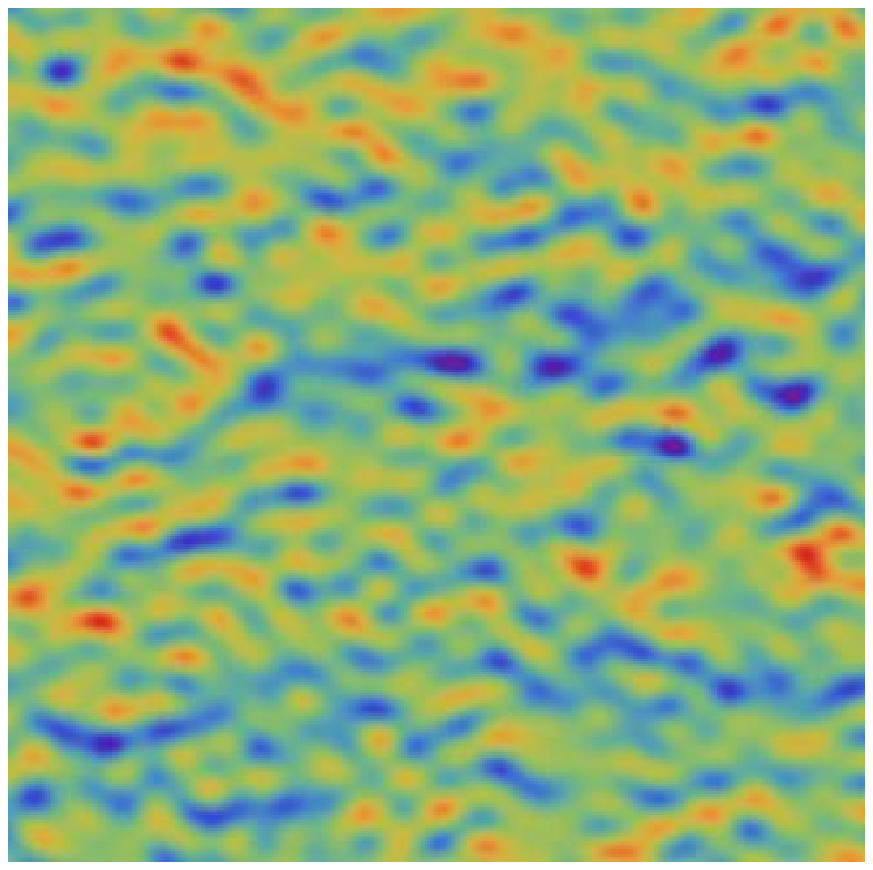}}}
{
\resizebox*{5cm}{!}{\includegraphics{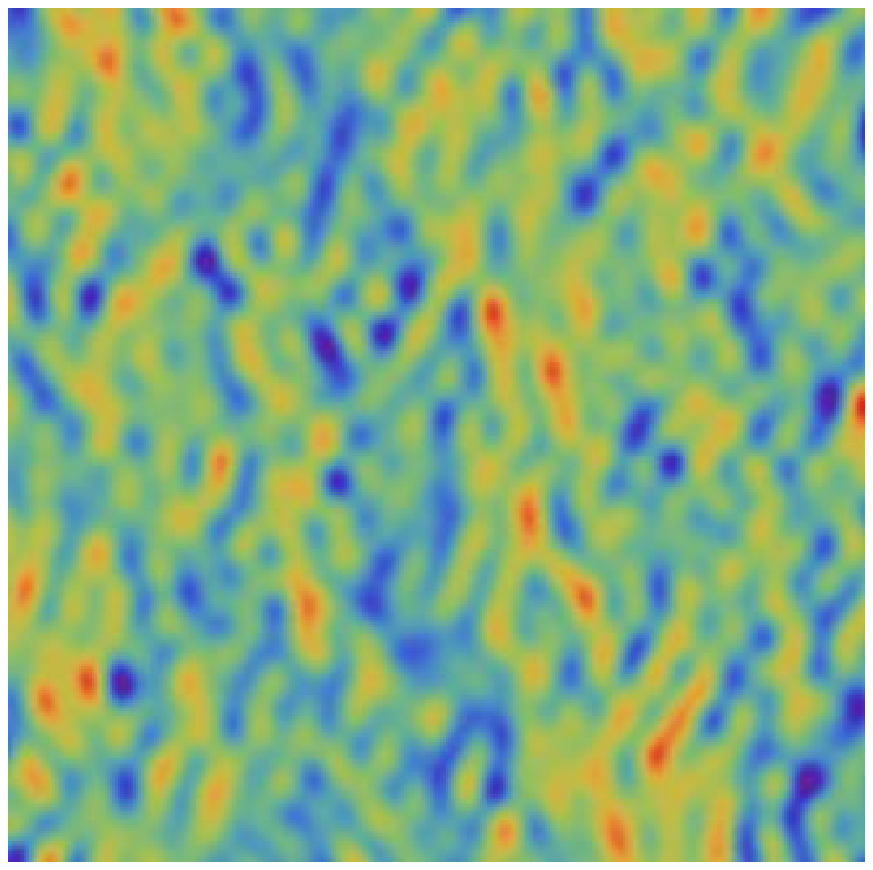}}}\\ %
}
\caption{%
Patterns in the CMB temperature perturbations which may arise from statistical
anisotropy in the CMB spectrum. The left panel shows
an isotropic signal, while the middle and right panels show patterns due to
statistical anisotropy along the horizontal and vertical direction 
respectively.}%
\label{stanis}
\end{minipage}
\end{center}
\end{figure}

A potentially clearer signal for the contribution of vector fields to the
primordial density perturbation is the appearance of statistical anisotropy in
the CMB temperature perturbations. Statistical anisotropy amounts to 
anisotropic patterns arising in the CMB designating special orientation. 
An example of how such patterns might look like is shown in Fig.~\ref{stanis}. 
One can think of them as similar to patterns in city planning produced by rows 
of identical houses. Statistical anisotropy in the CMB can arise due to 
anisotropic particle production during inflation \cite{stanis}. 
As mentioned, vector fields have three degrees of
freedom and can be analysed in three independent components. Each of these 
components undergoes particle production during inflation if the conformality 
of the vector field is suitably broken. However, the efficiency of the particle
production process may be different for each of these components and depends on
the mechanism which breaks the conformal invariance of the vector field. 

Statistical anisotropy is a new observable, which is inherently connected with
a vector field contribution to the density perturbations. So far there is no
observational detection of statistical anisotropic patterns in the CMB or in 
the distribution of structures in the Universe (e.g. rows of galaxies). 
However, the observational bounds are rather weak and allow the existence of 
statistical anisotropy at a level as high as 30\% \cite{obsanis}. The Planck 
satellite, which was launched in May of 
2009 by the European Space Agency, is expected to increase the CMB precision 
measurements by an order of magnitude and it is highly likely to observe 
statistical anisotropy. The main mission of the Planck satellite is to
observe non-Gaussian features in the predominantly Gaussian CMB temperature
perturbations. If a vector field contributes to the primordial density 
perturbation the non-Gaussianity in the CMB can also be anisotropic, in a 
direction which is correlated to the statistical anisotropy of the perturbation
spectrum \cite{anisnG}. This
is a smoking gun for the contribution of vector fields in the formation of 
structures in the Universe.

\subsection{Density perturbations and Galactic Magnetic Fields}

Before concluding let us discuss an interesting possible spin-off of the 
above ideas, which might connect the formation of structures in the Universe
with cosmic magnetism. 

\begin{figure}
\begin{center}
\begin{minipage}{100mm}
{\resizebox*{10cm}{!}{\includegraphics{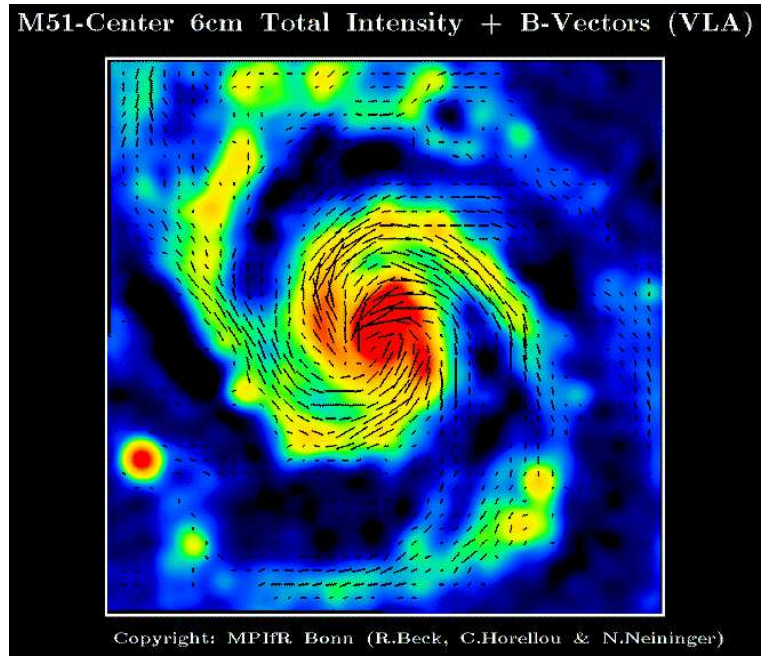}}}%
\caption{%
The observed magnetic field of the spiral galaxy M51. %
Contours denote the amplitude of the magnetic field while line-segments show
its direction. It can be clearly seen that the magnetic field follows the
spiral arms.
}%
\label{pmf}
\end{minipage}
\end{center}
\end{figure}

Observations suggest that the majority of galaxies carry magnetic fields of
equipartition strength: $B\sim 1\,\mu$G \cite{kron}, meaning that their energy 
density is comparable to the kinetic energy density of the galaxy. In spirals 
the magnetic field follows the spiral arms \cite{beck} (see for example 
Fig.~\ref{pmf}). This means that it is not frozen into the plasma
of the galactic disk\footnote{Were it frozen-in, it would have followed the 
differential rotation and not the density waves.}, which strongly suggests that
it is instead rearranged by a magnetic dynamo mechanism, similar to the one 
that operates in our Sun. The magnetic dynamo can also amplify an initial
magnetic field up to equipartition value (when dynamic backreaction stops 
further amplification) but it needs a non-zero seed field to feed on. For the
Sun the seed field is presumed to be the magnetic field of the Milky Way. For
galaxies at the time of formation the seed field needs to be at least as strong
as $B_{\rm seed}\sim 10^{-30}\,$G \cite{acd}
with a coherency scale $\ell\gsim 100\,$pc, 
corresponding to the largest turbulent eddy. The origin of this seed field 
remains elusive. 

\begin{figure}
\begin{center}
\begin{minipage}{100mm}
{\hspace{2cm}
\resizebox*{6cm}{!}{\includegraphics{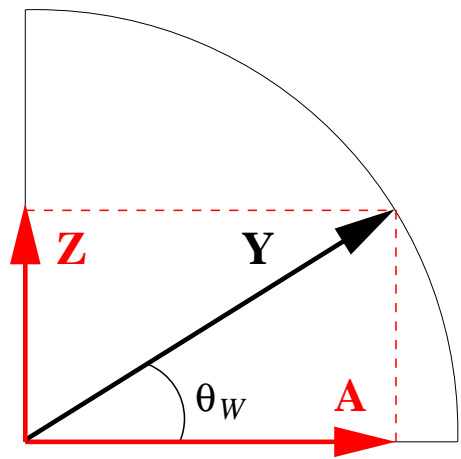}}}%
\caption{%
Projection of the hypercharge ($Y$) onto the photon ($A$) and the Z-boson ($Z$)
directions, through the Weinberg angle $\theta_W$, as it occurs at the breaking
of electroweak unification.
}%
\label{Ysplit}
\end{minipage}
\end{center}
\end{figure}

Suppose that the hypercharge vector boson field of the electroweak theory
(i.e. the unification of electromagnetism with the weak nuclear force)
has, through some mechanism, its conformality broken during inflation, so that
it obtains a flat superhorizon spectrum of perturbations. After inflation, the 
Universe cools down due to its expansion. At some moment, electroweak 
unification breaks down when the temperature of the Hot Big Bang reduces below 
$k_BT_{\rm ew}\sim 100\,$GeV, where $k_B$ is the Boltzmann constant. At the 
electroweak phase transition, the hypercharge field is projected onto the 
photon and the Z-boson directions, through the Weinberg angle as shown in 
Fig.~\ref{Ysplit}. The Z-boson condensate, being 
massive, rapidly oscillates a few tens of times before decaying to other 
lighter standard model particles. During these oscillations its superhorizon 
perturbation spectrum (which is due to the projection of the original
hypercharge spectrum) might affect the density perturbation in the Universe if
the density of the oscillating field is comparable to the density of the 
Universe at the time. The photon superhorizon perturbations (also due to the
hypercharge spectrum) give rise to a magnetic field with scale dependence 
$B(\ell)\propto 1/\ell$. Due to the high conductivity of the primordial 
plasma the magnetic field flux is conserved and the magnetic field survives
until the epoch of galaxy formation. It is a simple calculation to show that, 
provided the Z-boson has enough density to nearly dominate the Universe before 
its decay, the primordial magnetic field is strong and coherent enough to 
trigger the galactic dynamo and explain the magnetic fields of the galaxies
\cite{zcurv}. This scenario, if realised, may connect directly CMB observations
of statistical anisotropy with galactic magnetism.

\section{Summary and Conclusions}

According to state of the art cosmology, all structures in the Universe, 
including galaxies, stars, planets and ourselves, originated as quantum
fluctuations of suitable fields similar to the ones which we observe in the
lab with the Casimir experiment. These quantum fluctuations were transformed
into classical perturbations (giving rise to the primordial density 
perturbation) when stretched to sizes larger than the 
cosmological horizon (the range of causal correlations) during a brief period 
of superluminal expansion of space, called cosmic inflation. Inflation had to 
take place in the very early stages of the Universe history in order to solve 
one of the fundamental paradoxes of modern cosmology, the so-called horizon 
problem, i.e. the apparent uniformity of the Universe over distances which 
appear to be beyond causal contact. Inflation forces this uniformity onto the 
Universe by expanding an originally causally connected region of space to
size large enough to encompass the observable Universe at present. Inflation 
also creates the deviations from uniformity which are necessary for the 
formation of structures, by generating the primordial density perturbation,
which leads to structure formation through the process of gravitational 
instability. Recent precise observations of the perturbations of the 
temperature of CMB radiation, which reflect the primordial density 
perturbation, confirm both the particle production process through which
these perturbations were generated, and also cosmic inflation itself. These
observations are the earliest data at hand, since inflation has to take place
when the Universe is only a tiny fraction of a second old.

The precision of cosmological observations has reached the level which demands
theoretical model-building to become detailed and rigorous. Because of this 
and also in light of the forthcoming LHC findings it may be necessary to 
explore alternatives beyond the traditional scalar field hypothesis for the
realisation of the particle production process during inflation. One such 
possibility is considering vector boson fields. Massive vector fields
can indeed generate the primordial density perturbation without excessive 
large-scale anisotropy if they undergo coherent oscillations before they 
dominate the Universe. The scenario is characterised by a number of distinct
observational signatures. For example it may explain the observed weak 
large-scale anisotropy already found in the CMB (``Axis of Evil'') and it
can give rise to statistical anisotropy in the CMB temperature perturbations,
which amounts to direction dependent patterns that, at present, are allowed at
a level up to 30\% and may well be detected by the forthcoming observations
of the Planck satellite mission. Non-Gaussian features in the CMB 
perturbations may also be anisotropic in a manner correlated with the
statistical anisotropy in the CMB spectrum. This would be a smoking gun for
the contribution of a vector field in the density perturbation of the Universe.

\section*{Acknowledgements}

The author would like to thank the audiences of the 2008 Christmas Conference
of the Faculty of Science and Technology of Lancaster University and also the 
Physics Department of the University of Ioannina, for their feedback. The 
author also wants to thank his collaborators Mindaugas Kar\v{c}iauskas,
David. H. Lyth and Yeinzon Rodriguez-Garcia. This work is based on research 
funded (in part) by the Marie Curie Research and Training Network 
``UniverseNet'' (MRTN-CT-2006-035863) and by STFC Grant ST/G000549/1.

\end{document}